\documentclass[aps,prl,twocolumn,superscriptaddress]{revtex4}

\usepackage{graphicx}
\usepackage{amsmath}
\usepackage{color}

\begin{document}

\title{
Continuous `fish scale' planar electromagnetic meta-material}

\author{V. A. Fedotov}
\affiliation{EPSRC Nanophotonics Portfolio Centre, School of Physics
and Astronomy, University of Southampton, SO17 1BJ, UK}

\author{P. L. Mladyonov}
\affiliation{Institute of Radio Astronomy, National Academy of
Sciences of Ukraine, Kharkov, 61002, Ukraine}

\author{S. L. Prosvirnin}
\affiliation{Institute of Radio Astronomy, National Academy of
Sciences of Ukraine, Kharkov, 61002, Ukraine}

\author{N. I. Zheludev}
\email{n.i.zheludev@soton.ac.uk}
\homepage{www.nanophotonics.org.uk}
\affiliation{EPSRC Nanophotonics Portfolio Centre, School of
Physics and Astronomy, University of Southampton, SO17 1BJ, UK}

\date{\today}

\begin{abstract}
We report on a new type of continuous electromagnetic metal planar
meta-material, which resembles a `fish scale' structure. It is highly
transparent to electromagnetic radiation throughout a broad spectral
range apart from at one isolated wavelength. When the structure is
superimposed on a metallic mirror it becomes a good broadband
reflector everywhere apart from one wavelength where the reflectivity
is small. At this wavelength the reflected wave shows no phase change
with respect to the incident wave, thus resembling a reflection from
a hypothetical zero refractive index material, or `magnetic wall'. We
also discovered that the structure acts as a local field concentrator
and a resonant `amplifier' of losses in the underlying dielectric.
\end{abstract}

\maketitle

In optics, spectral selectivity in components such as filters,
beam-splitters and mirrors has traditionally been realized by
accurately engineering constructive and destructive multiple
interference of light in a stack of dielectric layers with of
thicknesses comparable to the wavelength. Spectral selectivity of
light transmission and reflection may also be found in regular
three-dimensional structures commonly know as photonic band-gap
crystals. However, another opportunity exists to achieve spectral
selectivity of transmission and reflection in a thin, essentially
sub-wavelength layer of material without engaging multiple beam
diffraction and interference. Wavelength sensitive transmission and
reflection of a thin layer may result from patterning the interface
on a sub-wavelength scale in a way that makes electromagnetic
excitation couple to the structure in a resonant fashion. The idea of
frequency selective planar structures has been investigated in the
microwave part of the spectrum for some time using arrays of
\emph{separated} holes in a metal screen or particles such as
crosses, snowflakes, tapers and split-ring resonators
\cite{vardaxoglou}. Similar research on the extraordinary
transmission of isotropic \cite{ebbesen} and anisotropic
\cite{elliott} arrays of holes in the optical part of the spectrum
has recently attracted a lot of attention. In this paper we point
out, however, that to achieve narrow spectral resonances
\emph{continuous} periodic structures may be used.

Here we report the first experiential results on a new type of
continuous planar meta-material structure - an equidistant array of
meandering metallic strips on a thin dielectric substrate producing a
pattern that resembles fish scales. In the past similar structures
have only been investigated theoretically \cite{wavy1}, \cite{wavy2}.
We show experimentally that the fish scale structure exhibits several
interesting properties when interacting with electromagnetic
radiation. First, it is highly transparent to electromagnetic
radiation across a wide spectral range apart from at one isolated
wavelength. Second, when such a structure is combined with
(superimposed on) a homogeneous metallic mirror with a sub-wavelength
dielectric spacing it becomes a good broadband reflector apart from
at an isolated wavelength where reflectivity is small. Third, at this
wavelength, there is no phase change of the reflected wave with
respect to the incident wave. This last property is particularly
unusual because on reflection from a conventional unstructured metal
or dielectric surface with refractive index $n > 1$, the electric
field of the wave accrues a phase reversal (phase shift of 180 deg.).
Indeed $E_\text{reflected} = -(n-1)/ (n+1) E_\text{incident}$, so
reflection without a phase change ($E_\text{reflected} =
E_\text{incident}$) implies an interface with a bulk homogeneous
medium of zero refractive index, $n = 0$. In microwave literature
this phenomenon is often referred to as reflection from a `magnetic
wall' \cite{sievenpiperMTT99}). Finally, the structure acts as a
local field concentrator and a \emph{resonant} multi-fold `amplifier'
of losses in the constitutive dielectric.

In the experiments reported here we used a fish scale structure
with a square translation cell of $15 \times 15$ $mm$ that was
etched from a $35$ $\mu m$ copper film on a fiberglass PCB
material of $1.5$ $mm$ thickness (see Fig. 1). The width of the
strips was $0.9$ $mm$. The overall size of the samples was
approximately $220 \times 220$ $mm$. We studied the reflection and
transmission of this structure in the $2-18$ $GHz$ spectral range.
The measurements were performed in an anechoic chamber using a
vector network analyzer (Agilent, model E8364B) and horn antennas
(Schwarzbeck~M.~E. model BBHA 9120 D). The structure does not
diffract electromagnetic radiation for frequencies lower than $20$
$GHz$ and therefore a single-wave regime was achieved in the
transmission and reflection experiments.

\begin{figure}[h]
\includegraphics[width=80mm]{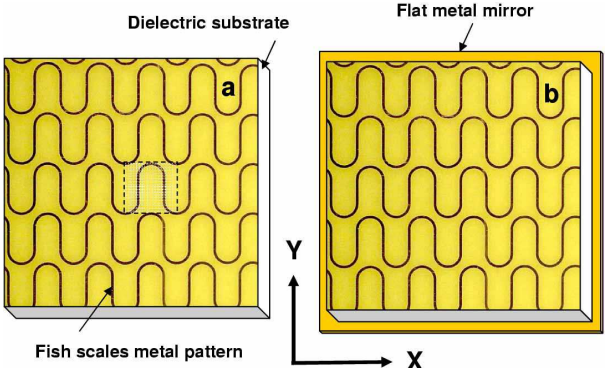}
\caption{a)~Fragments of the `fish scale' pattern of copper strips
on a dielectric substrate. The dashed line box indicates the
elementary translational cell of the structure. b)~The same
structure backed with a metal mirror.} \label{array}
\end{figure}

The results of a transmission experiment under normal incidence
conditions are presented in Fig. 2. The most dramatic spectral
selectivity is seen for the polarization state of incident light
parallel to the meandering metal strips (the $x$-polarization as
defined in Fig. 1). Here the transmission losses of the meander
structure are very small everywhere apart from in a sharp $-40$
$dB$ trough in transmission at $\nu_0 = 9.84$ $GHz$. The resonant
width at $+6$ $dB$ from the bottom of the trough is $\Delta\nu =
0.012$ $GHz$. For the orthogonal $y$-polarization (perpendicular
to the meandering strips) the transmission peak is shifted to
$6.18$ $GHz$ and the resonance width is about $1$ $GHz$.

\begin{figure}[h]
\includegraphics[width=80mm]{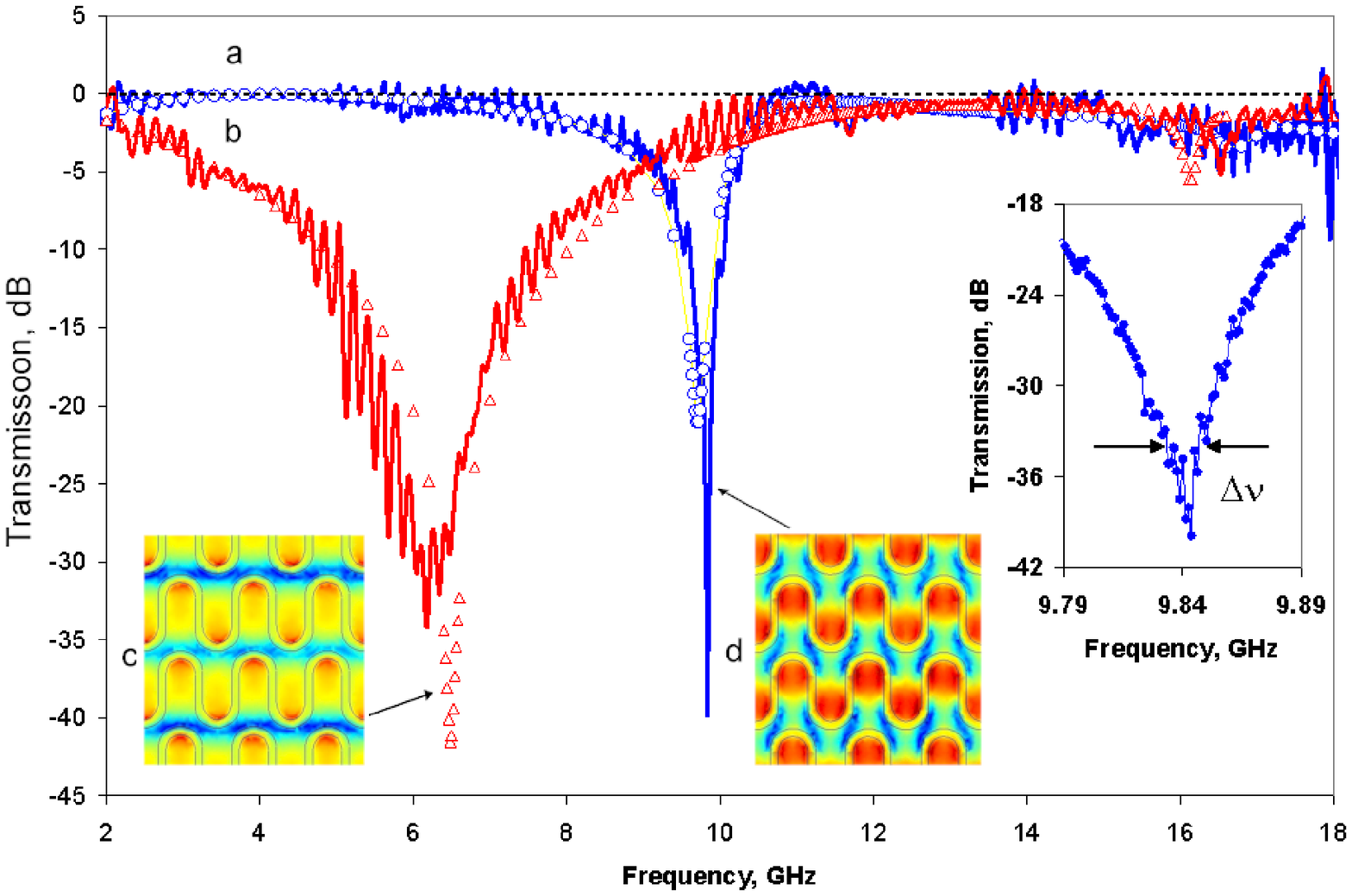}
\caption{Normal incidence transmission of the fish scale
meta-material presented in Fig. 1a for incident polarization along
(a) the $x$-direction (along the line of the meandering strips) and
(b) perpendicular to the strips. The color-coded maps (c) and (d)
show the instantaneous values of the electric field magnitude in the
plane of the structure at resonant conditions for the $y$- and
$x$-polarizations respectively (blue and red colors correspond
respectively to negative and positive values of the electric field).
The inset shows the resonant transmission trough taken at a high
spectral resolution. `o' and `$\triangle$' symbols denote the results
of theoretical calculations (method of moments) for $x$- and
$y$-polarized incident radiation respectively.} \label{dataT}
\end{figure}

Fig. 3 shows the results of a reflection experiment in which the
structure is combined with a homogeneous metallic copper sheet
mirror placed in contact with the opposite side of the dielectric
substrate supporting the fish scale structure, i.e. at a distance
of $1.5$ $mm$ from the fish scale structure. We studied reflection
under nearly normal incidence conditions with the incident wave
entering the sample at $6$ $deg.$ to the normal and with the plane
of incidence perpendicular to the line of the meandering strips.

Without a fish scale structure on top one would expect a metallic
sheet to be a very good reflector across the entire spectral range of
interest, and that the reflected wave would have the opposite phase
to the incident wave. What we found is that with an $x$-polarized
incident field the reflection losses are small everywhere apart from
in a $-40$ $dB$ trough at $8.77$ $GHz$. The resonant width at $+6$
$dB$ from the bottom of the trough is $\Delta\nu = 0.01$ $GHz$.
Refection for the perpendicular $y$-polarization is nearly perfect
everywhere apart from in a much smaller $-11$ $dB$ trough at $4.22$
$GHz$. For the $x$-polarization we observed that the reflected wave's
phase change has a strong dispersion and that the phase difference
between the incident and reflected wave passes zero in the proximity
of the the resonance, at $8.77$ $GHz$.

\begin{figure}[h]
\includegraphics[width=80mm]{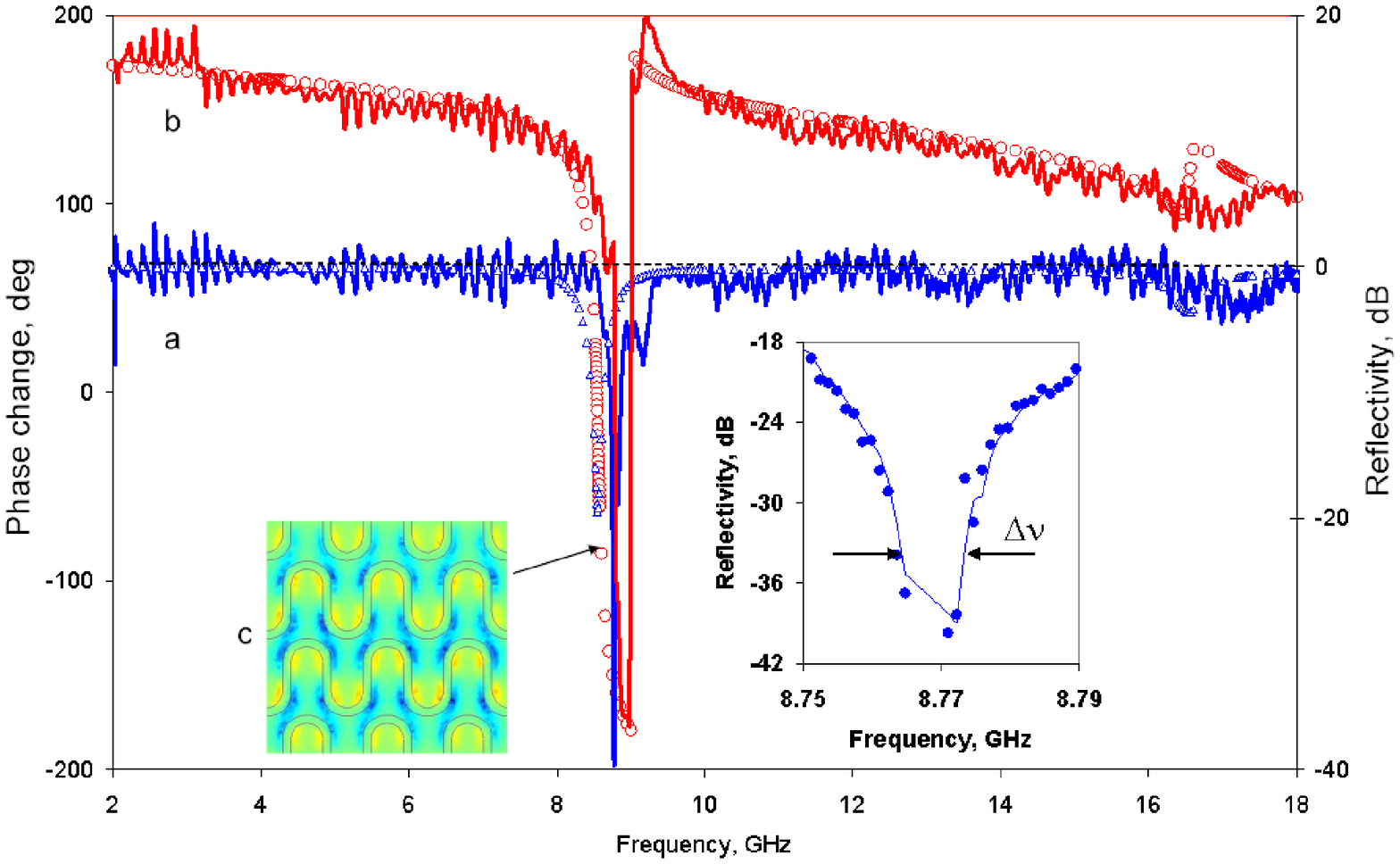}
\caption{Normal incidence reflection of $x$-polarized wave from the
fish scale meta-material backed with a flat copper mirror $mm$, as
presented in Fig. 1b: (a) Intensity of the reflected signal; (b)
Phase difference between the incident and reflected waves. The
color-coded map (c) shows the instantaneous values of the electric
field magnitude in the plane of the structure at the resonant
condition for the $x$-polarization (blue and red colors correspond
respectively to negative and positive values of the field). The inset
shows the resonant transmission trough at higher spectral resolution.
`$\triangle$' and `o' symbols denote the results of theoretical
calculations (method of momenta) for the intensity and phase of the
reflected wave respectively.} \label{dataM}
\end{figure}

The origin of the array's resonant features for $x$-polarized
incident radiation and their dependence on the characteristic
dimensions of the pattern becomes apparent if one considers the
structure as a sequence of straight-line strips periodically
loaded with short-circuit sections of length $S/4$ where $S$ is
the full length of the strip within the elementary translational
cell \cite{wavy1}. The wavelength of excitation in the structure
$\lambda_s \approx c/(\nu \sqrt{(\epsilon+1)/2})$, so the line
impedance has zero value at $\lambda_s= S/2$, corresponding to a
maximum in reflection and minimum in transmission. The substrate
material used to manufacture the fish scale structure had
$\text{Re} (\epsilon) \approx 4.5$ and an estimated reflection
resonance frequency of $9.38$ $GHz$, in very good agreement with
the measured value of $9.84$ $GHz$ for the minimum transmission
frequency (see Fig. 2). It is difficult to produce a simple and
accurate estimate for the resonant frequency for $y$-polarized
incident radiation, however this resonance seems to appear when
the wavelength of excitation in the strip is approximately equal
to $S$. Corresponding field distributions for $x$- and
$y$-polarized excitations in the plane of the structure are
illustrated in the insets to Fig. 2. They have been calculated
using the true three-dimensional finite-element method for solving
Maxwell's equations in the spectral presentation. The more
localized current distribution seen at resonance for the
$x$-polarization indicates weaker coupling between the array and
the free-space wave, and this manifests itself as a much narrower
resonance in comparison with the $y$-polarization.

For the fish scale structure on a metal-backed substrate the
following formula can be used to estimate the wavelength of the
excitation in the strips: $\lambda_g \approx (c/\nu)
[(\epsilon+1)/2+(\epsilon-1)/2/\sqrt{1+5h/w}]^{-1/2}$, where $2w$
is the width of the strip and $h$ is the substrate thickness. Now,
for the $x$-polarization the resonance occurs at a frequency where
$\lambda_g$ is close to half of the length $S$ of the strip inside
the elementary translation cell, i.e. $\lambda_g (\nu) \approx
S/2$. This estimation gives a resonant frequency of $8.77$ $GHz$
which coincides with the experimentally measured value. A
resonance for the $y$-polarization occurs when $\lambda_g (\nu)
\approx S$. The corresponding predicted resonant frequency is
$4.38$ $GHz$, which is in a good agreement with the measured value
of $4.22$ $GHz$.

We also modelled the far-field response of the fish-scale structure
using the method of moments \cite{wavy1}. This numerical method
involves solving the integral equation for the surface current in the
metal pattern by the field of the incident wave. This is followed by
calculations of scattered fields as a superposition of partial
spatial waves. The metal pattern is treated as a perfect conductor,
while the substrate is assumed to be a lossy dielectric. For
transmission through the fish scale structure the theoretical
calculations show very good agreement with the experimental results
(see Fig. 2). The numerical calculation also accurately describes all
of characteristic trends seen in the reflection experiment from the
combined fish scale/metal mirror structure, and predicts the dramatic
switching of the reflected signal phase at the resonance for the
$x$-polarization (see Fig. 3). Both the experimental results and
calculations show the phase of the reflected wave changing sign and
passing zero at about $8.77$ $GHz$, which is somewhat unusual
behavior. At zero phase change the reflected wave has the same sign
as the incident wave. In contrast, a wave reflected from a
conventional unstructured metal or dielectric surface has the
opposite sign to the incident wave. If the interface was loss-less,
which it could be in the case of an ideal metal and a non-absorbing
dielectric support, the zero phase change would correspond to
reflection from an interface with a bulk homogeneous medium of zero
refractive index, $n = 0$.

\begin{figure}[h]
\includegraphics[width=80mm]{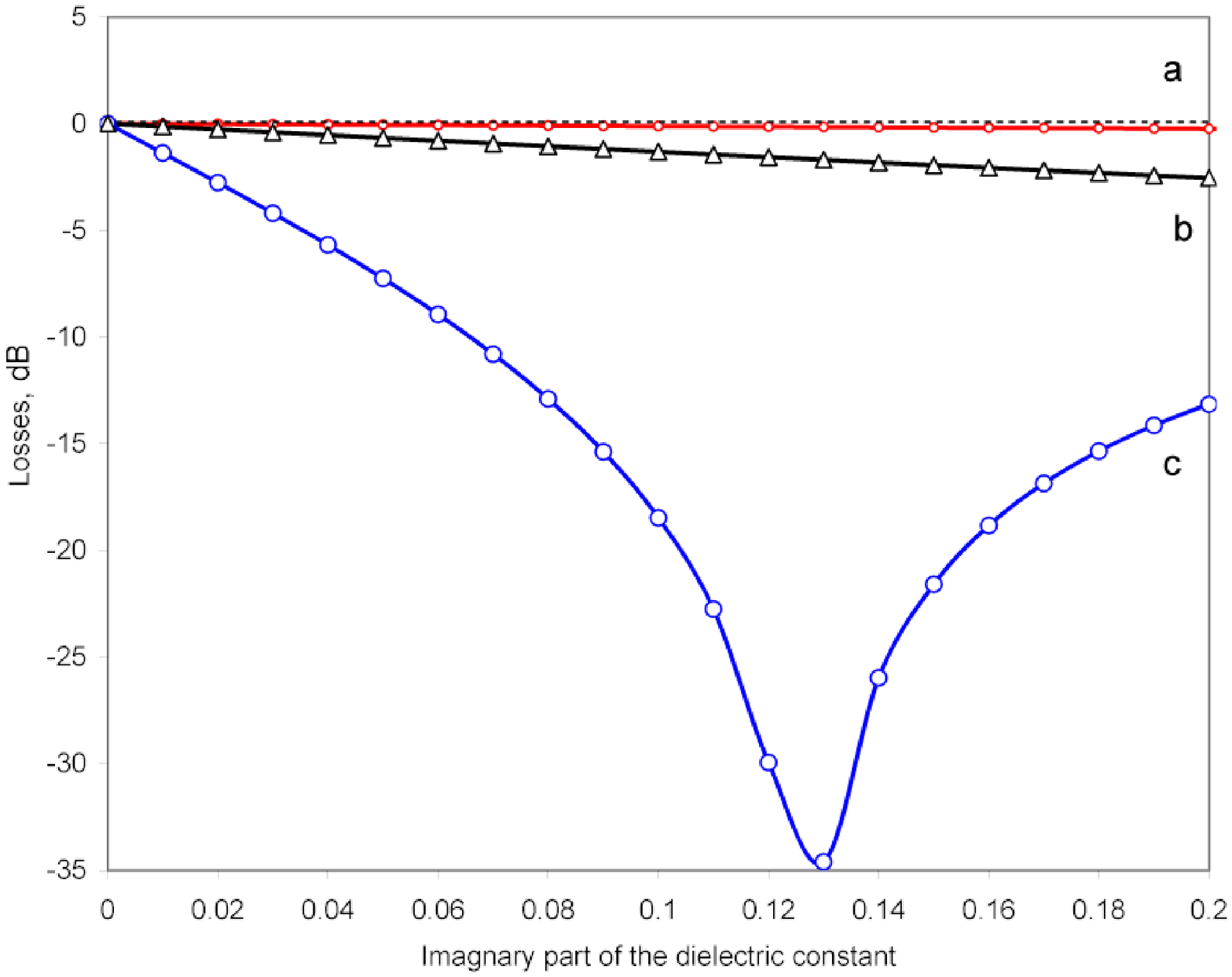}
\caption{Modelling losses in the fish scale meta-material as a
function of losses in the dielectric substrate for $x$-polarized
incident waves: (a) Reflection from a dielectric substrate on a
copper metallic plate (reference curve), (b) Attenuation due to
dielectric losses of the reflected signal (at resonance) from a fish
scale structure on a dielectric substrate, (c) Reflection from a fish
scale structure backed by a metallic mirror, with a dielectric layer
in between, at the $8.77$ $GHz$ resonance.} \label{dataM}
\end{figure}

Reflection from a flat mirror superimposed with the fish scale
structure shows another intriguing and potentially useful property -
`loss amplification'. Assuming that the strip structure and the
mirror are made out of ideal metal and that the dielectric is
loss-less, one would expect the reflected wave to have the same
amplitude as the incident wave. Thus, the trough in the reflected
signal, experimentally observed at the zero-phase frequency point,
should not appear for loss-less media. The only reason for such a
trough could be dissipative losses (other sources of losses such as
diffraction are not present in this structure because the patterning
period is smaller than the wavelength). However, in the microwave
part of the spectrum losses in copper are negligible, indeed
reflection losses for a free-standing copper mirror at $10$ $GHz$ are
about $2.8 \times 10^{-4}$. For a copper mirror with a dielectric
substrate in front of it the losses related to the double pass of the
wave through the substrate would only be a fraction of a decibel for
the microwave material used in our experiment. A simple loss
mechanism cannot by any means explain the high losses ($-40$ $dB$)
observed in the experiment at $8.77$ $GHz$ and their \emph{sharp
resonant character}. To explain strong resonant losses we calculated,
using the method of moments, the intensity of the reflected signal as
a function of losses in the substrate at the resonance (see Fig. 4).
With increasing losses in the dielectric substrate (i.e. increasing
value of the imaginary part of the substrate's permittivity)
reflection losses for a simple metallic mirror on the substrate are
small for a low-loss dielectric ($\text{Im} \{\epsilon\} < 0.2$)
(curve (a), Fig.~4). The attenuation of the reflected signal due to
dielectric losses can reach a few dB at resonance for the fish scale
structure \emph{without} a mirror behind it (curve (b) Fig.~4). In
contrast, the calculations predict strong resonant losses on
reflection from the fish scale structure when it is backed by a
metallic mirror on the low-loss substrate (curve (c), Fig.~4). These
calculations perfectly correlate with our experimental observations
(Fig. 3a) and show that at the zero-phase point the fish scale
structure dramatically `amplifies' the small losses of the
dielectric. This resonant amplification is related to the local field
factor i.e. the high concentration of field between the strips of the
fish scale design and the mirror plane.

The unusual loss amplification property of the fish scale structure
may be used for spectroscopy and characterization of materials where
the application of a fish scale structure to a sheet or film of
material could dramatically `amplify' its otherwise undetectable
losses. The weak sensitivity of a photo-detector outside its main
frequency band may be resonantly enhanced if a fish scale structure
is used as a light-harvesting material to enhance small tail
inter-band absorbtion. Here a photosensitive material should replace
the substrate dielectric of the structure. This approach may be
particularly efficient in semiconductor detectors and for increasing
the quantum efficiency of photomultipliers. A frequency-selective
detector may also be created with help of this structure. The x-y
anisotropy of the fish scale structure may be used in wave-plates
(retarders) for controlling the polarization state of transmitted and
reflected electromagnetic radiation. The ongoing development of
nano-fabrication techniques may well lead eventually to the use of
the fish scale structures in the optical part of spectrum.

\begin{acknowledgments}
The authors would like to acknowledge the assistance of M. V.
Bashevoy, S. Birtwell and K. F. MacDonald in CAD design and
preparation of the manuscript and the financial support of the
Engineering and Physical Sciences Research Council, UK.

\end{acknowledgments}

\end{document}